\documentclass[prl,aps,epsf,twocolumn,showpacs]{revtex4-1}
\usepackage[pdftex]{graphicx}
\usepackage{dcolumn}
\usepackage{bm}
\usepackage{epsfig}
\usepackage{latexsym}
\usepackage{amsmath}
\usepackage{amssymb}
\usepackage{color}
\usepackage{array}

\newcommand{\<}{\langle}
\newcommand{\e}{\varepsilon}

\renewcommand{\>}{\rangle}
\renewcommand{\(}{\left(}
\renewcommand{\)}{\right)}
\renewcommand{\[}{\left[}
\renewcommand{\]}{\right]}
\renewcommand{\v}[1]{\mathbf{#1}} 
\newcommand{\dslash}{d \hspace{-0.8ex}\rule[1.2ex]{0.8ex}{.1ex}}

\renewcommand{\d}{\partial}

\begin{document}
\title{Boundary-law scaling of entanglement entropy in diffusive metals}
\author{Andrew C. Potter}
\affiliation{Department of Physics, University of California, Berkeley, CA 94720, USA}

\begin{abstract}
Entanglement structure serves as a powerful way to characterize quantum many-body phases.  This is particularly so for gapless quantum liquids, where entanglement-based tools provide one of the only means to systematically characterize these complicated phases.  For example, the Fermi-surface structure of Fermi-liquids is revealed in entanglement entropy by a log-correction to the typical boundary-law scaling of simpler quantum ground-states.  In this paper, I analyze the entanglement structure of a disordered, but delocalized diffusive metal.  Using a combination of analytic arguments and numerical calculations, I show that, despite having the same number of extended gapless excitations as a clean Fermi-liquid, the diffusive metal exhibits only boundary-law entanglement scaling.  This result pinpoints the sharp Fermi-surface structure, rather than the finite density of gapless excitations, as the origin of the log-correction in the Fermi-liquid entanglement scaling.
\end{abstract}
\maketitle
\newpage

\noindent{\bf Introduction -}
Understanding the structure of entanglement provides a powerful basis for characterizing quantum many-body systems.  
In ground-states of local Hamiltonians in $d$-dimensions, the entanglement entropy of a sub region of size $L$ with the rest of the system typically scales like the boundary of the region $S(L)\sim L^{d-1}$ where $d$. By contrast excited states of thermal systems have extensive ``volume-law" entropy scaling, $S(L)\sim L^d$.  The limited entanglement structure of ground-states enables efficient simulations techniques like DMRG and higher-dimensional tensor-network analogs\cite{Schollwoeck,Cirac}.  Moreover, the distinction between boundary and volume law scaling allows a sharp definition of non-thermal many-body localized phases - for which even excited states exhibit boundary-law behavior\cite{Bauer_Nayak,SwingleMBL}.

Entanglement scaling can also be used to identify special properties of quantum ground-states.  For example gapped topologically ordered phases exhibit a universal sub-leading constant, $\gamma$, correction to boundary-law scaling: $S(L)\sim aL-\gamma+\dots$, whose value contains information about the topological excitations of the phase\cite{LevinWen,KitaevPreskill}.  

Whereas such gapped topological ground-states are quite well understood, the understanding of gapless quantum phases remains more rudimentary.  Here again, arguments based on entanglement have provided a powerful toolset for characterizing gapless phases.  For example, entanglement monotonicity properties such as the c-, F-, and a- theorems\cite{Zamolodchikov,CasiniHuerta_cThm,CasiniHuerta_FThm,Cardy,AThm} place strong constraints on possible phase diagrams\cite{TarunMonotonicity}, allowing one to non-perturbatively rule out various scenarios in contexts where there are limited or no applicable field-theoretic techniques.  

Entanglement scaling can also be used to directly distinguish different gapless phases in dimensions higher than one.  For example, Fermi-liquids with an extended momentum-space surface of gapless excitations are distinguished from scale invariant quantum critical points with gapless excitations only at a single point by a logarithmically diverging correction to the boundary-law scaling\cite{Klich,Li,Swingle}:
\begin{align} S_\text{FL}(L)\approx (k_FL)^{d-1}\log k_FL \end{align}
where $k_F$ is the Fermi wave vector.  There is also numerical evidence that this log-correction to boundary-law scaling survives when the quasi-particle nature of the electrons are destroyed resulting in a fractionalized non-Fermi liquid with a Fermi-surface worth of non-quasi-particle excitations\cite{Zhang}.  

Motivated by the fundamental role of entanglement based concepts in our understanding of higher-dimensional gapless phases, I analyze the entanglement entropy scaling of a diffusive metal.  Here, weak disorder ``erases" the Fermi-surface structure from (disorder-averaged) correlation functions on length-scales longer than the elastic mean-free path $\ell$.  However, despite the smearing of momentum space Fermi-surface structure, the system still contains a ``Fermi-surface's worth" of gapless and spatially extended excitations.  Hence, it is not a priori obvious whether entanglement entropy in a diffusive metal displays the log-violation to boundary-law scaling characteristic of a Fermi-surface.

In this work, I demonstrate that the log-correction to boundary-law scaling is absent in a diffusive metal, showing that the log-correction requires a sharp Fermi-surface structure\cite{Endnote:SharpFS}, and is not simply a property of having a finite density of extended gapless excitations.  

To establish this result, I employ a few complementary lines of attack.  First, based on the scaling of number fluctuations, I show that all Renyi entropies with Renyi index $\alpha\geq 2$ follow a boundary-law.  This strongly suggests boundary law behavior for the von-Neumann entropy, which is obtained from the Renyi entropies by analytically continuing $\alpha\rightarrow 1$.  This prediction is then verified by direct numerical computation for a non-interacting disordered Fermion system.  Next, I suggest an intuitive explanation for these results, based on a hydrodynamic description of a Fermi-liquid, that provides insight into the important physical distinction between the entanglement properties of diffusive and ballistic Fermi-liquids.  Lastly, I use a different but related example of a quadratically dispersing semimetal to sharpen the distinction between having a finite density of extended gapless excitations and having a Fermi surface.

\vspace{6pt}\noindent{\bf Number fluctuation based proof of boundary law Renyi entropies - } 
A non-local quantity, entanglement entropy cannot be completely reconstructed from a finite number of local observables.  Nevertheless, in free fermion systems, fluctuations of electron number, $N_A=\int_{r\in A}c^\dagger_rc_r$,  within a sub region $A$:
\begin{align} \sigma^2_{N_A} &= \<N_A^2\>-\<N_A\>^2 
= \int_{r,r'\in A} \<c^\dagger_rc_{r'}\>\<c_rc^\dagger_{r'}\>
\nonumber\\&=
\text{tr} M(1-M)\end{align}
tend to track the scaling of the entanglement between $A$ and its complement.  Here $M_{r,r'}=\<c^\dagger_rc_r'\>$ with $r,r'\in A$ is the matrix of two-body correlation functions.  

For example, the ground-state of  an ordinary insulator is a product state of exponentially well localized orbitals.  Then, uncertainty in the number of electrons in a region $A$ comes only from orbitals that straddle the boundary of $A$, and hence like entanglement exhibits boundary-law scaling.  As a second example, in a clean (ballistic) metal, like entanglement entropy, number fluctuations also exhibit a log-correction to boundary law scaling\cite{SwingleSenthil} (see also Appendix A):
\begin{align} \sigma^2_{N_A}|_\text{ballistic} \approx \alpha (k_FL)^{d-1}\log(k_FL)\end{align}
where $k_F$ is the Fermi wave-vector and $\alpha$ is a numerical constant, and $d$ is the number of spatial dimensions.

Generally, number fluctuations provide only a lower bound for the entanglement entropy:
\begin{align} S &= -\text{tr}\[M\log M+(1-M)\log(1-M)\] \label{eq:FreeFermionEntanglement}
\\\nonumber &= \sigma_N^2+
\sum_{n=2}^\infty \frac{1}{n}\text{tr}\[M(1-M)^n+(1-M)M^n\]\geq \sigma_N^2
\end{align} 
where $\text{tr}M = \sum_{r\in A}M_{r,r}$.
However, the reason the scaling of $\sigma_N^2$ closely tracks that of $S$ is that for the ground-state of a local quantum Hamiltonian, the majority of eigenvalues $\{\lambda_j\}$ of $M$ are essentially $0$ or $1$, and make negligible contribution to both entropy and number fluctuation.  The remaining sub-extensive fraction of non-negligible eigenvalues give roughly comparable $\mathcal{O}(1)$ contributions to both quantities. Equivalently, in the ground-state of a local Hamiltonian, the entanglement Hamiltonian: $H_\text{E} = \log \[\rho_A^{-1}-1\]$ has far fewer degrees of freedom than the sub region $A$ itself, where $\rho_A = \text{tr}_{A^c}\rho$ is the reduced density matrix of $A$.
Hence, both entanglement entropy and number fluctuations are essentially just ``counting" the number of non-negligible eigenvalues $\lambda_j$, i.e. the number of degrees of freedom in $H_E$, and tend to scale the same way.

In Appendix~A, I compute the disorder averaged number fluctuations, $\overline{\sigma^2_{N_A}}$, for a spherical sub region of radius $L$ in a three-dimensional diffusive metal and find that it follows the boundary-law scaling:
\begin{align} \overline{\sigma^2_{N_A}}|_\text{diffusive} \approx \beta\(k_FL\)^{d-1}\log(\gamma k_F\ell)\end{align}
where $\beta,\gamma$ are numerical constants and $\ell$ is the elastic mean-free path.  The absence of a log-correction in number fluctuations suggests that the entanglement entropy for a diffusive metal follows boundary law scaling, but does not directly rule out the possibility of a log-correction for the entanglement entropy.

Stronger evidence for boundary-law scaling of entanglement comes from considering the Renyi entropies:
\begin{align}
\mathcal{R}_\alpha &= \frac{1}{1-\alpha}\log\text{tr}\rho_A^\alpha 
\end{align}
These are closely related to the entanglement entropy, which is recovered in the limit of $\alpha\rightarrow 1$.
For a free fermion system, the Renyi entropies can be expressed in terms of the two-particle correlation matrix as\cite{Peschel,Song}: 
\begin{align}
\mathcal{R}_\alpha &=\frac{1}{1-\alpha}\text{tr}\log\[M^\alpha+(1-M)^\alpha\]
\end{align}
 I will now show that the number fluctuations bound the (disorder averaged) Renyi entropies for all real values of Renyi index $\alpha\geq 2$, i.e. that:
\begin{align} \overline{\mathcal{R}_\alpha} = f(\alpha)L^{d-1}+\dots 
\label{eq:RenyiBound}\end{align}
where $(\dots)$ indicate subleading in $L$ terms, and $f(\alpha)$ is some bounded function of Renyi index, $\alpha$.  It is natural to expect that $f(\alpha)$ has a unique analytic continuation to the full interval $\alpha\in [1,\infty]$.  In this case, the von-Neumann entanglement entropy is given by, $S=f(1)L^{d-1}+\dots$, and also satisfies boundary-law scaling.  These expectations will subsequently be checked numerically in the following section to ensure that no unexpected pathology arises in taking the limit of $\alpha\rightarrow 1$.  

To establish Eq.~\ref{eq:RenyiBound}, it is first useful to note  the eigenvalues of $M$ with unrestricted indices are just the occupation numbers of the energy eigenmodes of the system, which take values either $0$ or $1$.  Hence, by the Cauchy interlacing theorem, the eigenvalues of $M$ restricted to the sub region $A$ are also bounded between $0$ and $1$.  
\begin{widetext}
With this in mind, let us expand the $2^\text{nd}$ Renyi entropy as a power series in $M$ and $(1-M)$:
\begin{align} 
\mathcal{R}_2&= \sum_{n=1}^\infty\frac{2^n}{n} \text{tr}M^n(1-M)^n
= 2\text{tr}M(1-M)+\sum_{n=1}^\infty \frac{1}{n+1}\frac{2^{n+1}\text{tr}\[M^{n+1}(1-M)^{n+1}\]}{\text{tr} M(1-M)}
\leq  2\sigma_N^2+2\log(4/e)
\label{eq:R2Bound}
\end{align}
where in the last step I have used $0\leq\lambda_j\leq 1$ to bound $\sum_j \lambda_j^{n+1}(1-\lambda_j)^{n+1}\leq \frac{1}{4^n}\sum_j\lambda_j(1-\lambda_j)$.  
Moreover, $\mathcal{R}_2$ bounds all other Renyi entropies with $\alpha\geq 2$, since by Jensen's inequality $\frac{1}{1-\alpha}\sum_j\log\[\lambda_j^\alpha+(1-\lambda_j)^\alpha\]$ is a decreasing function of $\alpha$. Importantly, since these bounds hold for each given disorder realization, they are also true for disorder-averaged quantities.  
\end{widetext}
Since even gapped systems display boundary law scaling, we must also have $\mathcal{R}_\alpha\geq f_<(\alpha) L^{d-1}$.  Together with Eq.~\ref{eq:R2Bound}, this  establishes Eq.~\ref{eq:RenyiBound} with $f_<(\alpha)\leq f(\alpha)\leq \frac{2\sigma_N^2}{L^{d-1}}$.  Having established that all Renyi entropies with $\alpha\geq 2$ follow boundary-law scaling, I now turn to a numerical computation of the entanglement entropy.

\vspace{6pt}\noindent{\bf Numerical Results for the Spatial Scaling of Entanglement - }
To confirm the above analytic arguments, I numerically compute the entanglement entropy for non-interacting disordered metal from Eq.~\ref{eq:FreeFermionEntanglement}.
Specifically, I consider a square lattice tight-binding model  with dimensions $L_x\times L_y$ (in units of the lattice constant).  The uniform part of the Hamiltonian consists of nearest neighbor hopping with strength $t$ and chemical potential $\mu\approx 0$ (measured with respect to the band-center, i.e. near half-filling).  Disorder is modeled by a normally distributed random on-site potential with variance $W$, that is uncorrelated between different lattice sites.  Periodic boundary conditions are used to eliminate boundary effects.  The entanglement sub-region $A$ is chosen as a cylinder extending from $x=1$ to $x=L$, and which wraps around full system in the $y$-direction.

To completely evade localization and obtain a asymptotically diffusive metal on all length scales one would need to simulate either a fully three-dimensional system or a spin-orbit coupled two-dimensional system.  However, to avoid the computational difficulty of simulating such higher-dimensional systems, I instead simulate quasi-one dimensional systems of length $L_x$ much larger than width $L_y$.  In a strictly one dimensional system, there is no separation of scales between localization length $\xi$ and mean-free path $\ell$.  However for fixed $\ell$, $\xi$ grows with increasing strip width, $L_y$, and by choosing moderate values of $L_y$ one can obtain an order of magnitude or more separation between $\xi$ and $\ell$.

The mean-free path $\ell$ is estimated perturbatively in disorder strength as $\ell = 2\pi N(0)W^2$ where $N(0) = \frac{1}{\sqrt{L_y}}\sum_{j,\text{occ}}\frac{k_F^{(j)}}{2\pi v_F^{(j)}}$ is the density of states at the Fermi-level, and $k_F^{(j)}$ and $v_F^{(j)}$ are respectively the Fermi wave vector and velocity for the $j^\text{th}$ sub-band.  The localization length is obtained by numerically extracting the smallest Lyapunov exponent of the 1D transfer matrices , $\hat{T}_x$ of $1\times L_y$ slices of the 2D system at fixed $x$\cite{MacKinnon,Pichard}.  To avoid the exponential buildup of numerical inaccuracies, I employ a repeated orthogonalization procedure\cite{MacKinnon,Pichard} (see also Appendix B).  

\begin{figure}[bbb]
\includegraphics[width = \columnwidth]{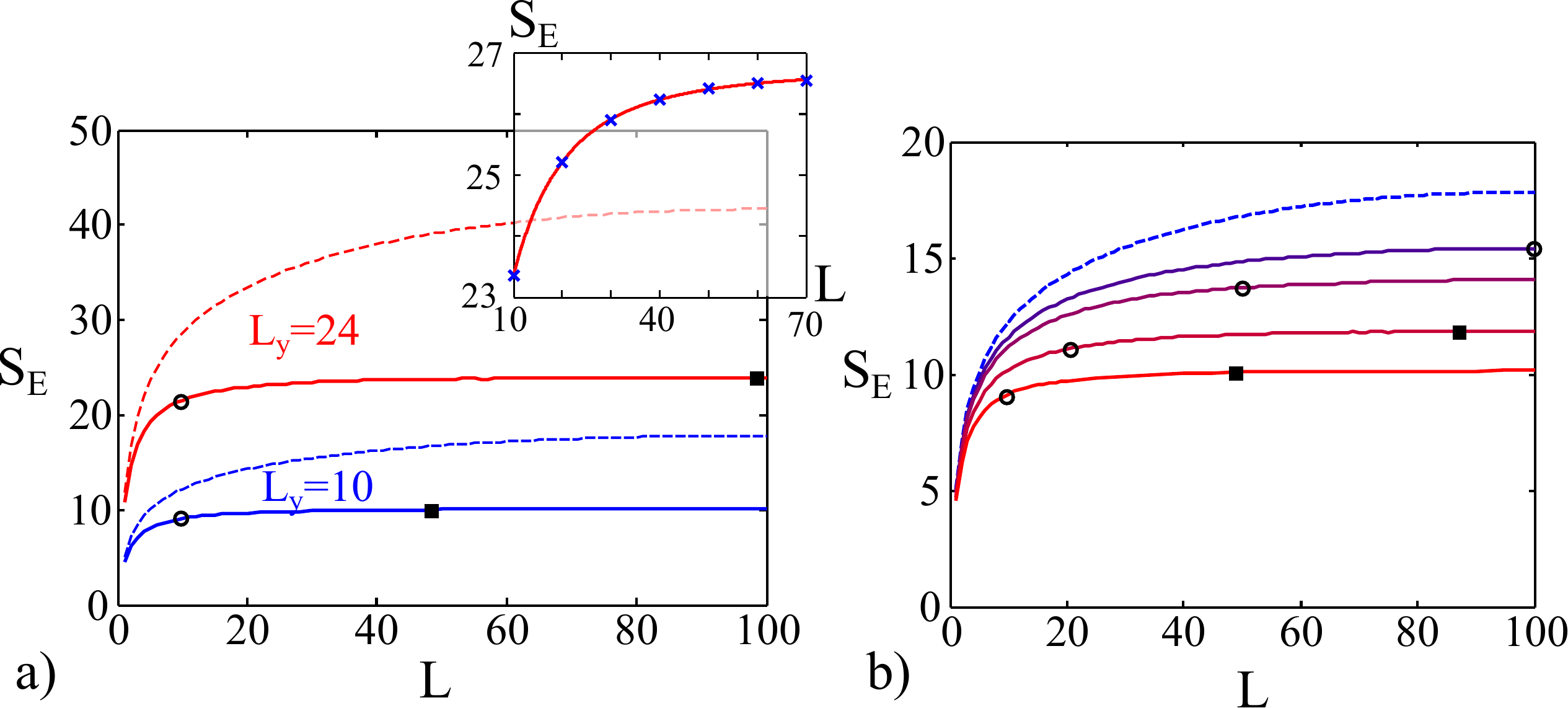}
\caption{Numerical results for the scaling of entanglement entropy $S_E$ for length $L$ segment of a disordered quasi-1D strip of dimension $L_x\times L_y$ near half-filling ($\mu=-0.1t$). Each curve is averaged over sufficiently many disorder realizations to reduce the error to $\lesssim 1\%$. Perturbatively computed mean free path, $\ell$, and numerically determined localization length $\xi$ are indicated by $\circ$, and $\blacksquare$ respectively. Dashed curves show clean $\ell=\infty$ results. Panel a) shows entanglement scaling for $L_x=200$, $L_y=10,24$.  The inset shows a fit to cut-off log-scaling of the form (inset) $\frac{c_\text{eff}}{3}\log\frac{1}{\sqrt{L^{-2}+\ell_*^{-2}}}+\text{const}$ for $L_x=150$, $L_y=24$, yielding $\ell_*\approx 15\pm 0.6$ and $c_\text{eff}\approx 11.3\pm 1.1$, consistent with the perturbatively calculated $\ell\approx 10$ and the number of occupied sub-bands (12). Panel (b) shows $S_E(L)$ for various disorder strengths corresponding to $\ell=10,20,50,100$ and $\infty$ for $L_x=200$, $L_y=10$.  The curves saturate at a scale consistent with the perturbatively computed $\ell$.  }
\label{fig:Numerics}
\vspace{-0.1in}
\end{figure}

Representative results are shown in Fig.~\ref{fig:Numerics}a for $\ell=10$, for $L_y=10$ ($\xi\approx 50$) and $L_y=24$ ($\xi\approx 100$).  One clearly sees that the initial log-growth of entanglement expected for a clean system is cutoff by the mean-free path $\ell$ such that $S(L)$ saturates to a constant for $L\gg \ell$ (see also Fig.~\ref{fig:Numerics}b for plots of $S(L)$ for various $\ell$ with $L_y=10$).  The analytic and numerical results show that the entanglement scales like $L_y\log L_x$ for $L_x\ll \ell$, and saturates to a constant for $L_x\gg \ell$.  The following function naturally interpolates between these two limits:
\begin{align} S(L)=\frac{c_\text{eff}}{3}\log\frac{1}{\sqrt{L^{-2}+\ell_*^{-2}}}+\text{const}
\label{eq:FitFunctionalForm}
\end{align}
As shown in the inset, the entanglement curves are well fit by the functional form in Eq.~\ref{eq:FitFunctionalForm} with $c_\text{eff}$ being essentially equal to the number of occupied sub-bands (i.e. is the central charge in the clean limit), and $\ell_*$ in reasonable agreement with the perturbatively estimated $\ell$.  Note that little change is seen as the system as $L$ passes through the localization length $\xi$, as $S(L)$ has already saturated at this point (note that in the localized regime $L\gg \xi$ $S(L)$ may not decrease below its asymptotic value in the diffusive regime due to the sub-additivity property of entanglement entropy\cite{TarunMBL}). 

In agreement with the analytic considerations, the direct numerical simulations clearly show an absence of log-corrections to area-law scaling of the ground-state entanglement entropy of a diffusive metal. The subsequent sections aim to provide additional physical intuition into these results.

\vspace{6pt}\noindent{\bf Intuition from a hydrodynamic description of the Fermi-liquid - }
Further insight into the above results can obtained from a ``hydrodynamic" description of the diffusive metal in terms of bosonic fields describing shape fluctuations of the Fermi-surface\cite{2DBosonization}.   by a ``hydrodynamic" description in terms of bosonic fields\cite{2DBosonization} describing shape-fluctuations of the Fermi-surface.  Consider an isotropic 3D Fermi surface.  The low-energy long-wavelength excitations of a clean Fermi-liquid can be approximately described by fluctuations of the Fermi-wave vector $\delta k_F(\hat{\Omega})(\v{r},t)$ that depend on both momentum-space angle $\hat\Omega$ on the Fermi-surface and which vary slowly in real space with characteristic scales $L\gg k_F^{-1}$.  The Fermi-surface disturbances can be equivalently parameterized by the extra density with momentum along solid angle $\hat{\Omega}$ on the Fermi-surface: 
\begin{align}\delta \rho(\hat{\Omega}) = \int_{|k|}\text{sgn}(k-k_F)c^\dagger_{\v{k}}c_{\v{k}}= \frac{k_F^2}{(2\pi)^3}\delta k_F(\hat{\Omega})\end{align}
The non-interacting Hamiltonian then reads: $H_0=\frac{16\pi^4v_F}{k_F^4}\int d\hat{\Omega}\rho_{\hat{\Omega}}^2$.  Assuming rotational invariance for simplicity, and introducing angular Harmonics,  $\rho_{l,m}=\int d\hat{\Omega}Y_{l,m}(\hat{\Omega})\rho(\hat{\Omega})$ gives: 
\begin{align} H_0 = \frac{16\pi^4v_F}{k_F^4}\sum_{l,m}\rho_{l,m}(\hat{\Omega})^2\end{align}

In the presence of random impurities, the overall density mode $\rho_{0,0}$ propagates diffusively:
\begin{align}\overline{\<\rho_{0,0}(t,\v{r})\rho_{0,0}(0,0)\>}\sim e^{-r^2/D|t|}\end{align}
whereas the remaining modes (for example $\rho_{1,1}$, which corresponds to net electrical current along the $z$-direction) decay exponentially in both space and time 
\begin{align}
\overline{\<\rho_{l,m}(t,\v{r})\rho_{l,m}(0,0)\>}\sim e^{-r/\ell_{l,m}}e^{-t/\tau_{l,m}}
\end{align}
Here $\overline{(\dots)}$ indicates average over disorder, $D$ is the diffusion constant, and $\tau_{l,m}\approx \ell_{l,m}/v_F$ are disorder induced relaxation times, which are equal for point-like disorder but in general can depend on $(l,m)$.

Therefore, in a loose sense, that the disorder averaged system contains only a single propagating hydrodynamic mode with dynamical exponent $z=2$, which in dimensions $d>1$ contributes only area law contribution to entanglement (plus possible subleading corrections).  By contrast, the clean system contains an infinite number of ballistically propagating hydrodynamic modes each contributing a logarithmically diverging entanglement entropy that add up to give $L^{d-1}\log L$ entanglement scaling.  The hydrodynamic picture provides an intuitive, albeit slightly hand-waving, understanding for the fact that a diffusive metal has only area-law contributions to entanglement entropy.  In particular, whereas the above methods apply only for free fermion systems, the hydrodynamic picture explained here is unchanged by electron-electron interactions, suggesting that interacting diffusive metals will also satisfy boundary-law entanglement scaling.

\vspace{6pt}\noindent{\bf Distinction Between Finite Density of Extended States and a Fermi-Surface - } 
In a diffusive metal, I have shown that disorder removes log-corrections to boundary law entanglement scaling on the same length scale, $\ell$, as it smears the Fermi-surface structure.  This strongly suggests that such log-corrections are a unique entanglement signature of a sharp Fermi-surface, and are not produced simply by having a large number of extended gapless excitations.  In this section, I provide further support for this picture by noting a second example of a metallic system with finite density of extended gapless excitations but no Fermi-surface, and which exhibits boundary law scaling with no log-correction.

Consider a (clean) two-dimensional semimetal, such as bilayer graphene, with a quadratic band touching (QBT) described by the one-body Hamiltonian:
\begin{align} H_\text{QBT}(p_x,p_y)=\frac{1}{2m^*}\begin{pmatrix} 0 & p_-^2\\ p_+^2 & 0 \end{pmatrix} \end{align}
where $p_\pm = p_x\pm ip_y$, and $m^*$ is the effective band mass.  For concreteness, consider a system with periodic boundary conditions, infinite extent in the x-direction, and large, but finite size $L_y$ in the y-direction. Let us compute the entanglement between the cylindrical sub region: $A=\{0<|x|<L,\forall~ y\}$ and its complement.  Due to the $y$-direction translation symmetry of $A$, $H_\text{QBT}$ decomposes into $\mathcal{O}(L_y)$ transverse sub-channels with fixed $p_y \in \{\frac{2\pi n}{L_y}\}$ 
%
and dispersion: $E_n(p_x)=\sqrt{\(\frac{p_x^2}{2m^*}\)^2+\Delta_n^2}$, where $\Delta_n = \frac{\pi^2n^2}{2m^*L_y^2}$. The entanglement of $A$ with the rest of the system is just the sum of the corresponding 1D entanglement entropies for each sub-band:
\begin{align} S(L_x,L_y) &= \sum_n S_\text{1D}(L_x,\Delta_n) \nonumber\\
&\approx \frac{\sqrt{2m^*}}{\pi}L_y\int \frac{d\Delta }{\sqrt{\Delta}} S_\text{1D}(L_x,\Delta)
\label{eq:QBTEntanglement}
\end{align}
While the $p_y=0$ channel is gapless, it actually just decomposes into completely filled and empty bands that happen to touch at $p_x=0$, and hence gives only boundary-law contributions to $S$.  Moreover, all of the sub-channels with non-zero $p_y$ are gapped one-dimensional systems and also should give only boundary-law contributions.  To shore up this argument, one needs to ensure that there are no anomalous contributions from vicinity of small  but non-zero $p_y$.  For this purpose, it is sufficient to focus on the regime of small $\Delta$, where all length scales, $L_x$ and $\frac{1}{\sqrt{m^*\Delta}}$, are much longer than the lattice scale, $a$, and scaling arguments (see also Ref.~\cite{SwingleSenthil})  may be applied. 

Specifically, consider the contribution to Eq.~\ref{eq:QBTEntanglement} from $\Delta\in[0,\Delta_0]$ where $\Delta_0$ is any suitable cutoff satisfying $\frac{1}{m^*a^2}\gg\Delta_0 \gg \frac{1}{m^*L_x^2}$. 
For simplicity let us choose units such that $2m^*/\pi^2=1$ to drop the length-independent prefactor in Eq.~\ref{eq:QBTEntanglement}. Then, at low energies, scale invariance of the theory with dynamical exponent $z=2$ requires that: $S_\text{1D}(L_x,\Delta) = S_\text{1D}^\text{(reg)}+\mathcal{S}(\Delta L^2)$ where $ S_\text{1D}^\text{(reg)}$ is a non-universal non-singular piece from short-distance physics which may contain at most area law contributions, and $\mathcal{S}(x)$ is a universal scaling function containing long-wavelength contributions. Then the derivative of the long-wavelength contributions to $S$ from $\Delta \ll \Delta_0$ is:
\begin{align}\frac{\d S_\text{LW}(L_x,L_y)}{\d L_x} &\approx L_y\int d\Delta~ \mathcal{S}'(x) \approx \frac{L_y}{L_x^2}\int dx ~x\mathcal{S}'(x)
\end{align}
Since $\mathcal{S}(x)$ is the entanglement entropy of a region of size $x$ the gapped 1D system with Hamiltonian $H_n$ (with length measured in units $\xi_\Delta$) it must follow an boundary law, i.e. $\mathcal{S}(x)\approx \text{const}+\mathcal{O}\(\frac{1}{x}\)$.  Hence the integrand $x\mathcal{S}'(x)\leq \mathcal{O}\(\frac{1}{x}\)$, and $\frac{\d S}{\d L_x}\leq \frac{L_y}{L_x^2}\int \frac{dx}{x} \approx \frac{L_y \sqrt{\Delta_0}\log L_x}{L_x^2}$ where I have used $x_\text{max}=\sqrt{\Delta_0}L_x$ as a short distance cutoff.  Since the $\frac{1}{L_y}\frac{\d S_\text{LW}}{\d L_y}$ vanishes in the limit of $L_x\rightarrow \infty$, $S_\text{LW}$ contains at most area-law contributions $\sim \text{const}\times L_y$.

This demonstrates that, like the diffusive metal, the quadratically dispersing semimetals follow boundary-law entanglement scaling, lending further support to idea that an extended Fermi-surface rather than just finite density of states is required to produce a log-violation of boundary-law scaling.

\vspace{6pt}\noindent{\bf Discussion -}
In this work, I have established that diffusive metals obey boundary-law scaling of entanglement entropy, in contrast to clean metals with a Fermi-surface, which exhibit a characteristic log-violation of boundary law scaling.  While the calculations undergone here are valid for non-interacting fermions, the hydrodynamic picture explained above suggests that these results to hold also for interacting diffusive metals. 

In conjunction with general monotonicity requirements for entanglement entropy, these results could possibly be used to constrain the entanglement scaling and possible phase diagrams for phases that are continuously connected to a diffusive metal.  To this end, a potentially interesting subject for future work would be to understand where there are any universal structure to the sub-leading corrections, e.g. analogous to the subleading constant piece, $F$, of 2D conformal systems (see however \cite{SwingleNonMonotonicity}), and if so whether they satisfy analogous monotonicity properties under renormalization group flows.

\vspace{12pt}
\noindent\emph{Acknowledgements}
I would like to thank Ashvin Vishwanath, Romain Vasseur, Siddharth Parameswaran, and especially Brian Swingle for helpful conversations.  This work was supported by the Gordon and Betty Moore Foundation.

\newpage
\appendix
\onecolumngrid
\setlength{\leftskip}{0.5in}
\setlength{\rightskip}{0.7in}

\section{Appendix A - Computation of number fluctuations in ballistic and diffusive metals}
In this appendix I compute number fluctuations in a sub region of size $L$ for a ballistic metal and a diffusive metal.
Consider a spherical sub region, $A$, of radius $L$.  The variance of electron number in $A$ is given by:
\begin{align} \sigma^2_{N_A} &= \<N_A^2\>-\<N_A\>^2 
= \int_{r,r'\in A} \<c^\dagger_rc_{r'}\>\<c_rc^\dagger_{r'}\>  
\nonumber\\
&= \int_{r,r'\in A} \int_{k,\omega,k',\omega'} e^{ik\cdot(r-r')}e^{ik'\cdot(r'-r)}A_{k,\omega}A_{k',\omega'}n_F(\omega)(1-n_F(\omega'))
\label{eq:NumberFluctuations}
\end{align}
where $A_{k,\omega} = \frac{g^R(k,\omega)-g^A(k,\omega)}{2\pi i}$ , and $g^{R/A}$ are the retarded/advanced one-electron Green functions.
\begin{align} \sigma^2_{N_A}  = \int_{q,\omega,\omega'}&n_F(\omega)(1-n_F(\omega'))\left|\int_{r<L}e^{iq\cdot r}\right|^2\times
\nonumber\\
&\times \(\frac{1}{2\pi^2}\)\int_k\(g^R(\omega,k+q)g^A(\omega',k)+g^A(\omega,k+q)g^R(\omega',k)\)
\end{align}
where I have used that $\int_k g^{R}g^R = 0 = \int_k g^Ag^A$.

\begin{figure}[htb]
\vspace{24pt}
\includegraphics[width = 6in]{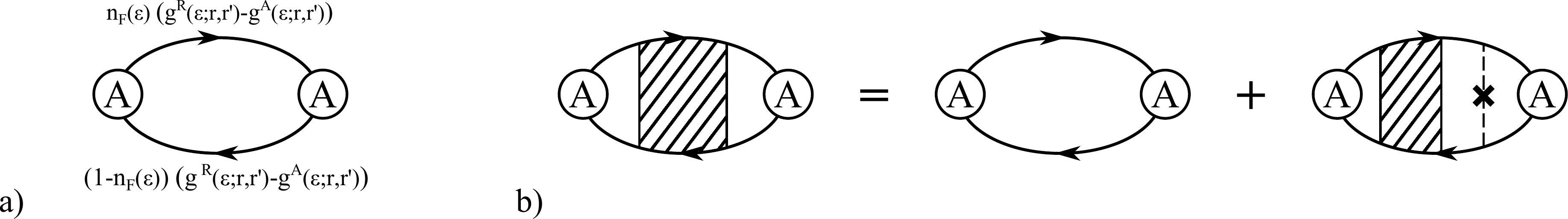}
\caption{Diagrammatic representation number fluctuations for a ballistic clean metal (a) and the corresponding disorder averaged quantity in a diffusive metal (b).  Solid lines indicate bare Green functions (a) or disorder averaged Green functions (b).  Encircled $A$ vertices indicate integration over points within the region $A:\{\v{r}:|\v{r}|<L\}$.  The dashed line with an $\times$ in the middle indicates scattering from an impurity.}
\label{fig:NumberFluctuations}
\vspace{12pt}
\end{figure}

\subsection{A.1 - Number fluctuations in a clean ballistic metal}
In a clean metal: $g^{R/A} = \(\omega-\e_k\pm i0^+\)^{-1}$.  Linearizing $\e_k$ near the Fermi-surface gives:
\begin{align} \int_k g^R(\omega,k+q)g^A(\omega',k) 
&\approx \int \dslash\hat{\Omega}_k~ \nu(0)\int d\xi\frac{1}{\omega-\xi-v_F\cdot q+i0^+}\frac{1}{\omega'-\xi-i0^+}
\nonumber\\&=
\nu(0) \int \dslash\hat{\Omega}_k~\frac{2\pi i}{\omega-\omega'+v_F\cdot q+i0^+}
\end{align}
where $\nu(0)\approx \frac{k_F^2}{2\pi v_F}$ is the density of states at the Fermi-surface, and $k_F,v_F$ are respectively the Fermi wave vector and velocity.

Then I find:
\begin{align}
&\int_k\(g^R(\omega,k+q)g^A(\omega',k)+g^A(\omega,k+q)g^R(\omega',k)\)
=
\int_k\(g^R(\omega,k+q)g^A(\omega',k)+c.c\)
\nonumber\\
&=
4\pi^2\nu(0)\int\dslash\hat{\Omega}_k\delta\(\omega-\omega'+v_F\cdot q\) = 4\pi^2\nu(0)\int_{-1}^1\frac{d\mu}{2}\delta\(\omega-\omega'+v_Fq\mu\)
\nonumber\\
&= \frac{2\pi^2\nu(0)}{v_Fq}\theta\(1-\frac{|\omega-\omega'|}{v_Fq}\)
\end{align}
which, upon performing the frequency integrals becomes:
\begin{align}\int_{\omega,\omega'}n_F(\omega)(1-n_F(\omega')) \theta\(1-\frac{|\omega-\omega'|}{v_Fq}\)&= \frac{1}{4\pi^2}\int_{\omega'=0}^{v_Fq}\int_{\omega=-v_Fq}^{\omega'-v_Fq} = \frac{1}{4\pi^2}\int_{\omega'=0}^{v_Fq}\omega' = \frac{1}{8\pi^2}(v_Fq)^2
\end{align}

Inserting these expressions into Eq.~\ref{eq:NumberFluctuations} gives:
\begin{align} \sigma^2_{N_A} = \int_q \left|\int_{r<L}e^{iq\cdot r}\right|^2 \frac{\nu(0)}{4}v_Fq
\end{align}
The Fourier-transformed projection into the sub region evaluates to:
\begin{align}\int_{r<L}d^3re^{iq\cdot r} = \int_{r<L}r^2dr \frac{\sin{qr}}{qr} = \frac{\sin qL-qL\cos qL}{q^3}\end{align}
Assembling these expressions gives:
\begin{align} \sigma^2_{N_A} = \frac{\nu(0)}{4}\int_q \(\frac{\sin qL-qL\cos qL}{q^3}\)^2v_Fq
\end{align}

In the limit of large sub region size, $L\rightarrow \infty$, the number fluctuations read:
\begin{align} \sigma^2_{N_A}\approx \frac{v_F\nu(0)}{8\pi}~ L^2\int dq \frac{\cos^2qL}{q} \approx \alpha(k_FL)^2\log k_FL\end{align}
where $\alpha$ is a numerical constant, and I have used the Fermi-momentum as a short-distance cutoff, remembering that our expressions were obtained by linearizing the electron dispersion near the Fermi-surface. Note that, like the entanglement entropy, the number fluctuations scale like $(k_FL)^2\log(k_FL)$.  The extra factor of $\log(k_FL)$ comes from a log-divergence at small wave-vectors $q$, which is cut-off only by the finite size of the sub region under consideration.  In the next section, I perform the analogous disorder-averaged calculation for the diffusive metal and find that there is no such long-wavelength singularity in that case, and correspondingly no $\log$ correction to the boundary law scaling of number fluctuations.

\subsection{A.2 - Number fluctuations in a disordered diffusive metal}
In the presence of impurity scattering, the disorder averaged Green's functions develop an elastic lifetime $\tau$:
\begin{align} \overline{g}^{R/A}=\frac{1}{\omega-\e_k\pm\frac{i}{2\tau}\pm i0^+}
\end{align}
To obtain the disorder averaged scaling of number fluctuations, however, one must disorder average the entire Eq.~\ref{eq:NumberFluctuations} (shown diagrammatically in Fig.~\ref{fig:NumberFluctuations}b).

For simplicity, let us assume a density $n_\text{imp}$ of impurities with a short-range potential approximated by a delta-function of strength $V_0$ (related to the elastic scattering rate by $\tau^{-1}=2\pi \nu(0)V_0^2n_\text{imp}$).
In the limit $\e_F\tau\gg 1$ the disorder averaged product of Green functions $\Pi(\Omega,q) = \int_k\overline{g^R(\omega+\Omega,k+q)g^A(\omega,k)}$, can be written as a geometric series: $\Pi = \Pi_0+V_0^2n_\text{imp}\Pi = \frac{\Pi_0}{1-V_0^2n_\text{imp}\Pi_0}$ in the product of disorder averaged Green functions:
\begin{align} \Pi_0(\Omega,q) &= \int_k \overline{g}^R(\omega+\Omega,k+q)\overline{g}^A(\omega,k) = \nu(0)\int d\xi \int_{-1}^1\frac{d\mu}{2}\frac{1}{\omega+\Omega-\xi-v_F q\mu+\frac{i}{2\tau}}\frac{1}{\omega-\xi-\frac{i}{2\tau}}
\nonumber\\
&=\nu(0)\int_{-1}^1\frac{d\mu}{2}\frac{2\pi i}{\Omega+v_Fq\mu+\frac{i}{\tau}}\approx \frac{2\pi i\nu(0)}{\Omega+iDq^2+\frac{i}{\tau}}
\end{align}
Where, the last line is valid in the limit of $\Omega,Dq^2\ll \tau$.

Summing the geometric series gives the familiar diffusive form of $\Pi$:
\begin{align} \Pi(\Omega,q) = \Pi_0+V_0^2n_\text{imp}\Pi = \frac{\Pi_0}{1-V_0^2n_\text{imp}\Pi_0} = \frac{2\pi i\nu(0)}{\Omega+iDq^2}\end{align}
 
The real part of $\Pi$ is then:
\begin{align} \Pi+c.c. = \frac{2\pi\nu(0)Dq^2}{\(\omega-\omega'\)^2+\(Dq^2\)^2} \end{align}
Inserting this expression into Eq.~\ref{eq:NumberFluctuations} gives:
\begin{align} \overline{\sigma^2_{N_A}}\approx \int_{\omega<0,\omega'>0}\int_q\(\frac{\sin qL-qL\cos qL}{q^3}\)^2\Pi(\omega-\omega',q)
\end{align}

For the clean metal, the log-correction to area law came from a long-wavelength (i.e. infra-red or IR) divergence from the small $q$, which was cut off by the finite size of the sub region when $q\lesssim \frac{1}{L}$. Let us examine the IR behavior coming from $\omega,Dq^2\ll \tau^{-1}$, in the limit of $L\rightarrow \infty$:
\begin{align} \overline{\sigma^2_{N_A}}\large{|}_{\frac{1}{\ell}\gg q\gg \frac{1}{L}\text{ Regime}}& \approx \nu(0)\int_{\omega<0,\omega'>0}\int q^2dq \frac{L^2}{q^4}\frac{Dq^2}{(\omega'-\omega)^2+(Dq^2)^2}
\nonumber\\ &\approx
\nu(0)DL^2\int dq \int_0^\infty d\omega \frac{1}{Dq^2}\(\frac{\pi}{2}-\tan^{-1}\(\frac{\omega}{Dq^2}\)\)
\nonumber\\ &\approx 
\nu(0)DL^2\int dq \[\int_{\omega \ll Dq^2}\frac{\pi}{2Dq^2}+\int_{\omega\gg Dq^2}\frac{1}{\omega}\]
\approx 
\nu(0)DL^2\int_0^{\ell^{-1}} dq \[\log\(\frac{1}{\tau Dq^2}\)\]
\nonumber\\ &\approx
\frac{\nu(0)D}{\ell}L^2
\end{align}
Here I have introduced the elastic mean-free path $\ell = v_F\tau$, and $\ell^{-1}$ and $\tau^{-1}$ serve as short wavelength and high frequency cutoffs to the momentum and frequency integrals respectively since the diffusive form of $\Pi$ applies only for $\omega,Dq^2\ll \tau^{-1}$.  For the ballistic regime, $\omega, Dq^2\gg \tau^{-1}$, disorder is unimportant and the contributions from this regime coincide with those of a clean-metal computed in the previous section (but with IR cutoff $\ell$).  

Piecing together the contributions from the diffusive and ballistic regimes, gives number fluctuations scaling as:
\begin{align} \overline{\sigma^2_{N_A}} \approx L^2\[c_1\frac{\nu(0)D}{\ell} +c_2k_F^2\log{k_F\ell}\]\approx \beta\(k_FL\)^2\log(\gamma k_F\ell)\end{align}
where $\beta,\gamma,c_{1,2}$ are numerical constants.  Hence, the number fluctuations of the diffusive metal do not have a log-correction to area law scaling.

%
%

\section{Appendix B. Algorithm for Extracting Localization Length}
The Lyapunov exponents are given by the log of the eigenvalues of the limiting matrix:
\begin{align} \hat{T}_\infty=\lim_{N\rightarrow \infty} \(\prod_{j=1}^N\hat{T}_x\)^{1/N} 
\label{AppEq:LimitingTMatrix}
\end{align}
The smallest Lyapunov exponent corresponds to the inverse localization length. However, as written, Eq.~\ref{AppEq:LimitingTMatrix} is unsuitable for direct numerical computation, since repeated products of transfer matrices are dominated by the largest Lyapunov exponent, resulting in rapid (exponential in $N$) accumulation of numerical errors for the localization length.  

I employ the standard trick\cite{MacKinnon,Pichard} of periodic orthogonalization to avoid this numerical instability. This approach begins by choosing a block size of $N\sim 10$ sites and computing: $\hat{T}_{1}=\prod_{x=1}^{N}\hat{T}_x$.  A QR-decomposition of $\hat{T}^{(1)} = Q^{(1)}R^{(1)}$ is then performed, where $R$ is an upper right-triangular matrix whose diagonal entries contain the eigenvalues of $\hat{T}^{(1)}$ (sorted from highest to lowest) and $Q^{(1)}$ is an orthogonal matrix.  The Lyapunov exponents are then computed iteratively as follows: define $\hat{T}^{(j+1)}=\prod_{x= Nj+1}^{(N+1)j}Q^{(j)}$.  Here $Q^{(j)}$ is the orthogonal matrix from the QR composition: $\hat{T}^{(j)}=Q^{(j)}R^{(j)}$.  In this procedure, as $j$ is increased, the $k^\text{th}$ column of $Q^{(j)}$ is increasingly oriented along the eigenvector of $\hat{T}_\infty$ with $k^\text{th}$ largest eigenvalue.  This approximate orthogonalization reduces the random contamination of small Lyapunov exponents by larger ones in subsequent steps.

The Lyapunov exponents are then given by $\lambda_n=\lim_{j\rightarrow \infty}\frac{1}{Nj}\sum_j\log (R^{(j)})_{n,n}$.  The localization length is then the inverse of the smallest Lyapunov exponent $\xi=\frac{1}{\min_n\lambda_n}$.  In practice, a large number $\gtrsim 10^5$ of iterations are required to determine localization length with accuracy of $\approx 1\%$.


\begin{thebibliography}{10}
\bibitem{Endnote:SharpFS}
Here I define ``sharp" Fermi surface as having a continuous momentum space surface, for example the $2k_F$ components of density associated with Friedel oscillations, at which certain correlation functions exhibit singularities.  This definition includes non-Fermi liquids that lack coherent quasi-particles but which may nevertheless have a ``sharp" Fermi surface structure in the sense defined here, and which also exhibit log-corrections to boundary law scaling\cite{Zhang}.
\end{thebibliography}
\end{document}